# Mass Flow Analysis of SARS-CoV-2 for quantified COVID-19 Risk Analysis

Gjalt Huppes and Ruben Huele, 14 October 2020, revised 29 October 2020

## *Abstract*

How may exposure risks to SARS-CoV-2 be assessed quantitatively? The material metabolism approach of Industrial Ecology can be applied to the mass flows of these virions by their numbers, as a key step in the analysis of the current pandemic. Several transmission routes of SARS-2 from emission by a person to exposure of another person have been modelled and quantified. Start is a COVID-19 illness progression model specifying rising emissions by an infected person: the human virion factory. The first route covers closed spaces, with an emission, concentration, and decay model quantifying exposure. A next set of routes covers person-to-person contacts mostly in open spaces, modelling the spatial distribution of exhales towards inhalation. These models also cover incidental exposures, like coughs and sneezes, and exposure through objects. Routes through animal contacts, excrements, and food, have not been quantified.

Potential exposures differ by six orders of magnitude. Closed rooms, even with reasonable (VR 2) to good (VR 5) ventilation, constitute the major exposure risks. Close person-to-person contacts of longer duration create two orders of magnitude lower exposure risks. Open spaces may create risks an order of magnitude lower again. Burst of larger droplets may cause a common cold but not viral pneumonia as the virions in such droplets cannot reach the alveoli. Fomites have not shown viable viruses in hospitals, let alone infections. Infection by animals might be possible, as by cats and ferrets kept as pets. These results indicate priority domains for individual and collective measures. All Supporting Information can be downloaded from www.ScienceforStrategies.com/SARS-Covid .

The wide divergence in outcomes indicates robustness to most modelling and data improvements, hardly leading to major changes in relative exposure potentials. However, models and data can substantially be improved.

**Institutions:**
Gjalt Huppes: Department Industrial Ecology, CML, Leiden University, Leiden, Netherlands
Ruben Huele: Department Industrial Ecology, CML, Leiden University, Leiden, Netherlands

**Corresponding Author:**
Gjalt Huppes
Oudezijds Achterburgwal 119C
NL1012 DE Amsterdam
Netherlands
huppes.cml@gmail.com

## 1	INTRODUCTION: Mass Flows of SARS-CoronaVirus-2
### 1.1	Industrial Ecology for SARS-CoV-2 and COVID-19

In environmental sciences the analysis of toxic substances has a base structure of primary production and elimination at source; processes in the environment with transport, transformation, concentration, and breakdown; and finally, exposure by humans, and by other creatures. The prime tool for this analysis is material flow analysis, for large volumes by mass and for smaller entities by their numbers. The new SARS-CoV-2 viruses (SARS-2 for short) are discrete entities, the size of Ultra Fine Particles (UFPs) at around 0.1μm hence are measured by their numbers. In the human factory they are produced, re-used, and controlled. The part remaining is emitted in different forms: as virion UFPs; as virion clusters of PM 2.5 size, the droplet nuclei remaining after mini-droplet evaporation; contained in larger watery and mucal droplets; and as solid waste, as stool excrement. These constitute the human factory emissions of an infected person, see flows 1 to 4 in Figure 1. Several factories may constitute a heavily contaminated industrial site.

After emission there are different routes through the environment for the SARS-2 virions, ultimately still viable virions entering a person that potentially becomes a new COVID-19 patient: a new factory. As with toxic substances, these different routes towards possible exposure lead to different forms, concentrations, and volumes of contagion, with different options for immission reduction. The risks resulting are even more divergent. SARS-2 virions, like fine asbestos particles, are quite harmless on the skin but very harmful if arriving in the alveoli or intestines. Quantifications are prerequisite for assessing risks, as basis for the design of prevention measures. The overall SARS-2 production should be non-circular and decreasing. This quantified Industrial Ecology approach is virtually lacking in current SARS-2 to COVID-19 discussions.

There is ample literature on the medical, virologic, and epidemiological aspects of SARS-2 and the Covid-19 illness it causes. Disparate approaches, with overlapping concepts, make the discussion and analysis difficult. Airborne emissions range from breathing, speaking, singing, sneezing and coughing, with immissions by inhalation of droplets of different sizes, airborne droplet nuclei and single virions, and acquiring viruses from physical contacts with larger droplets and contaminated objects brought to mouth and nose, including a route starting from excrements. For all these single elements in the Covid trajectories abundant literature is available. However, the full trajectories are lacking, let alone their combined and comparative analysis. That gap is filled by this paper. Of course, this is done in a preliminary way, as a framework to be improved upon as new information and better models become available.

### 1.2  SARS-2 Virion Flow Scheme

A person ill with COVID-19 pneumonia has virion production primarily in his alveoli, over 50m$^2$, with internal secondary infections mostly in the small intestines, together resulting in four types of emission, nrs 1 to 4, see the overall framework in Figure 1. There are direct airborne emissions (1) and emissions of very small virions containing droplets (2). As these mini-droplets evaporate directly at exhalation, except in extremely humid circumstances, both are airborne emissions. Their shares are broadly disputed. Virions in larger droplets are the third type of emission. Data on virion density in them is mostly lacking with first approaches in the older review by (Gralton, Tovey, McLaws, & Rawlinson, 2011) and for SARS-2 in (Vuorinen et al., 2020). The second main human production site is in the small intestines, covering around 35m$^2$. Viable virions are found in excrements (4), emitted in diverse ways. Production and emission are modelled in Section 2 and quantified in Section 3 see the supporting information in www.ScienceforStrategies.com/SARS-Covid .

The routes through the environment are modelled in Section 3, with results in Section 4. SARS-2 virions require ACE2 receptor cells for replication so only humans are sources, with possibly a minor role for animals like ferrets and cats. Also, SARS-virions break down quite fast. So, the routes towards exposure are relatively simple, only close to the sources. Main quantified attention is given first to closed spaces where concentrations can build up. Next, the direct person-to-person routes are quantified: the exhalations in breathing, speaking, and singing and the bursts of coughs and sneezes, inhaled by a person. A more semi-quantitative approach is developed for direct inhalation of larger droplets, on very short distance, and on intake to nose and mouth by contact exposure as with fomites.

In Section 5 the quantified results of all routes to exposure are brought together, ordered as to the amount of potential virions intake. Results are discussed in Section 6. Technical measures and policy measures for prevention, a complex subject, are not treated. Priority domains come up clearly however, with some conclusions in Section 7.

## 2  METHODS 1: Emission Model
### 2.1  From primary production to emission

The ill person is a human factory of the SARS-2 viruses, with SARS-2 production sites and emission sites, and with removal mechanisms reducing emissions, the factory analogue. There are two independent approaches for the quantification of emissions. One is based on modelling of the primary production and all the steps towards emission, building on empirical knowledge of the parts and on mass balance principles. The second approach measures emissions directly, as flows from the factory chimneys, the nose and mouth, and in solid wastes as stool excrements. Both methods are followed here, with model data aligned to empirical data recently becoming available.

Production facilities are concentrated in a few places in the body, only there where cells have an ACE2 receptor. SARS-2 (and SARS-1) can only replicate in such cells, see (Hamming et al., 2004), for a full survey and with different emphasis (Bourgonje et al., 2020). Prime production is in the lung's alveoli. Their combined surface is over 50m$^2$ and much more at inhalation (Fröhlich, Mercuri, Wu, & Salar-Behzadi, 2016). At Day 21 in the model, see Table 1, 2% of the 480 million alveoli (Ochs et al., 2004) are infected, over 1m$^2$ of infected cells, rising to 20m$^2$ at day 25. Such an infection on normal skin would make a person seriously ill. In the lungs, with only two cells thick surface to blood, it is serious SARS-2 pneumonia. The second main production location is in the small intestines, with a surface of at least 35m$^2$. Much smaller amounts of these cells are in the airways, especially secretory and ciliated cells in the nose (but not so in the bronchi); in sweat glands in toes and hands; and in many small veins and arteries.

**Figure 1 Primary SARS-2 production and emission, environmental transformation and immission**

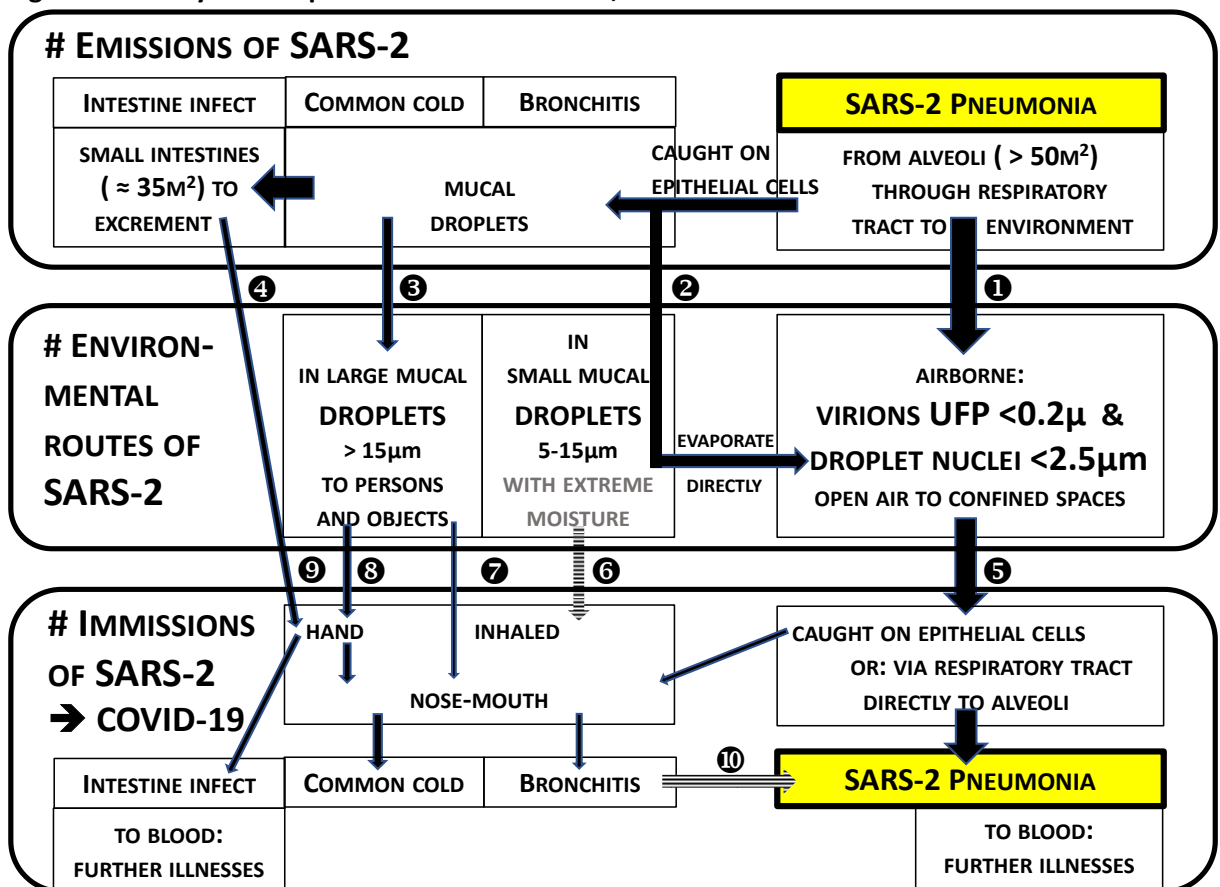

Legenda to Figure 1:

| EMISSIONS | IMMISSIONS | PNEUMONIA |
|---|---|---|
| 1. Airborne exhalation of virions | 5. Inhaled airborne viruses | 10. Indirect infection through airways infection: Not documented |
| 2. Evaporated mini-droplets, airborne | 6. Mini droplets inhaled | |
| 3. Large droplets as from sneezing and coughing | 7. Large droplets inhaled | |
| | 8. Remains of large droplets brought to nose-mouth | |
| 4. Stool containing viruses | 9. Stool remains to nose-mouth | |

The surfactant-wetted virions produced leave the alveoli through the alveolar tubes and next enter the airways at the smallest bronchioli. At normal breathing many of these tubes collapse at end of exhalation, their mucal and watery walls taking in particles. With inhalation they enter the alveoli as minidroplets leaving the filled alveoli at next exhalation. They mostly are below 1µm, see (Bake, Larsson, Ljungkvist, Ljungström, & Olin, 2019). At deep inhalation and exhalation, the tubes don't close and ultra-small particles such as virions can move directly to the airways, and out of them, emitting. That mini droplet forming process is described in detail in (Malashenko, Tsuda, & Haber, 2009).

Contrary to SARS-1, around 80% of the SARS-2 virions produced are not viable (Ogando et al., 2020) Table 6b. Normal defense mechanisms, from trapping to destructing, take out a major part of outgoing viable SARS-2 virions, and very similar of incoming virions, see (Fokkens & Scheeren, 2000) on involved mechanisms. Non-specific defense mechanisms may already have some relation with SARS-2, as possibly based on previous non-SARS corona common cold infections (Mateus et al., 2020). Specific defense mechanisms develop later after infection only, then halting the otherwise exponential growth. Part of the mini-droplets caught in the mucus of the epithelial cells remains viable. It is mainly swallowed and may infect the also sensitive small intestines. Part may be coughed or sneezed out as large droplets, falling, and one part may exit as the second type of mini-droplets, airborne first but then as droplet nuclei at usual direct evaporation after exhalation. The share of mini-droplets from mucus is not known but must be a small share of virions in mucus, as most mucus is swallowed, and large droplets dominate sneezes and, less so, coughs, see (Scharfman, Techet, Bush, & Bourouiba, 2016) Fig. 5 & 6.

The descriptions here refer to a basically healthy person infected. However, impaired health may change the shares. People breathing very superficially, because of lung illnesses or obesity, will not have fresh air deep into their lungs to be exhaled; a larger share of viable virions than 15% might remain inside for reinfection. Also impaired lung functioning reduces the effectiveness of clearing particles from the airways, especially ultrafine particles (Brown, Zeman, & Bennett, 2002) to which individual virions belong. The overall destruction rate, set at 96%, will then be lower, the advance of illness faster and then emissions rising faster.

**Figure 2. Primary production and respiratory emissions of a SARS-2 infected person**

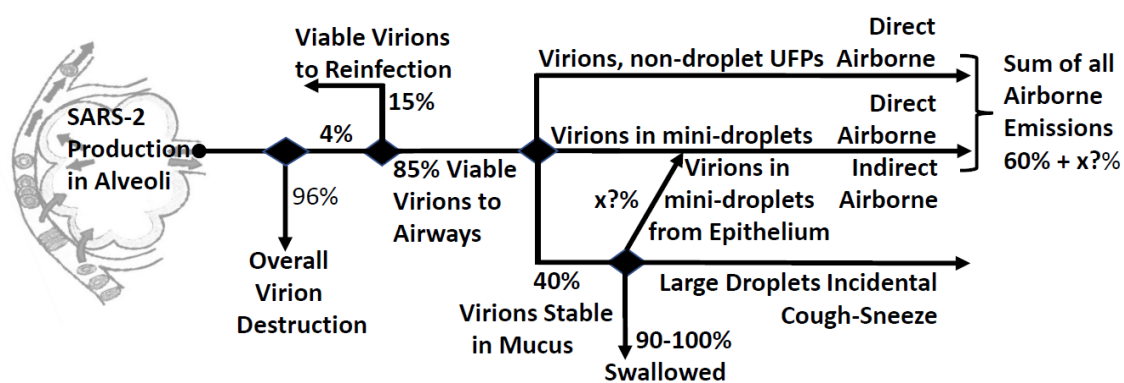

## 2.2   Exponentially expanding production modelled

Production of SARS-2 starts after effective infection. Next, the newly produced virions, the part not destroyed by defenses or emitted, enter other alveoli of that person. This leads to exponential growth unless specific defenses against SARS-2 come into play, halting the exponential growth. There is some discussion on how many virions should be taken in for effective infection, with airborne virions passing the alveolar tube to an alveolus. The theoretical minimum is 1 while a minimum of 100 is mentioned in (Mittal, Ni, & Seo, 2020). The model here assumes a starting number of 24 alveoli infected by at least one virion. As most virions entering the respiratory tract are wiped out, 96% is assumed, the starting viral intake is set at 600, see Table 1. The Reference Person for emission is at Day 17, exhaling 74321 virions per hour. Starting with only 1 infected alveolus would add 5 days to the route to illness, starting already with 1000 cells infected would reduce the time to illness with 6 days.

In the first week or so the infected person will be asymptomatic, then develops symptoms, and at day 21 hospital admission becomes due, moving to the IC Unit a few days later if own defenses and medicines don't bend the curve downward. This roughly covers the period from infection to PCR measurable infection, to symptoms and to serious illness if occurring, see the survey by (Gaythorpe et al., 2020), Figure 3 for a general impression, and the estimated days in (Petersen et al., 2020). The corresponding growth rate per day of 1.8 (doubling time 28 hours, compound growth 2.5% per hour) seems a reasonable first estimate. It will be slower for young and healthy persons and faster for persons with compromised basic defenses. The model indicates that heavy exposure leads to faster and more serious illness.

Real life measurements of numbers of viable viruses exhaled require complex apparatus developed only a few year ago, see (Pan et al., 2016). PCRs cannot measure the viability of the collect nor really the concentration. In a modern high quality hospital room with two Covid-19 patients and one of them with a positive nose swab, (Lednicky et al., 2020) found a concentration of viable SARS-2 virions in air of 6 to 74 TCID50/L) corresponding to around 4000 to 52 000 virions As the ventilation speed and air mixing is not known this number cannot be linked to the number of exhaled virions but is compatible with outcomes of (Ma et al., 2020). In a sample of just ill to just-hospitalized patients, Ma et al. measured between 6000 and 600 000 viable virions per hour, with wide variations between persons and for the same person at different times. Taking the logarithmic mean of their results, the 60 000 corresponds roughly to our modelled Day 17 number, while the 6000 links to Day 13 and 600 000 to Day 21. This is second corroboration of the speed of a person's infection development in the model.

This virion production location of SARS-2 is very different from that of influenza. There replication is in the respiratory tract, not in the alveoli. Viral cold, sore throat and bronchitis are prime flu illnesses, much more severe there than in COVID-19. However, influenza breaks down the respiratory epithelial defense mechanisms against intruders. Bacterial infection to the alveoli may result, with possibly a deadly bacterial pneumonia following, indirectly. With SARS-2, it is very different: The bronchi are hardly affected, but a viral pneumonia is induced directly, in the ACE2 receptor cells of the alveoli, see (Hamming et al., 2004).

## 3   RESULTS 1: Emissions
### 3.1   Distribution of SARS-2 virions over respiratory outflows

The total share of production coming out in mucus - swallowed or sneezed and coughed out - is set somewhat arbitrarily. Let us assume a distribution of the virions going upwards in the airways as 40% caught stably in mucus and 60% as mini droplets with virions and free-floating virions. The mini-droplets and virions are exhaled, together covering 20% of the primary virions production in the alveoli. The number of free-floating viruses emitted nearly doubles every day, proportional with primary production. At a certain stage, the virus-specific defenses must come up or the patients will die. With exponential growth, 96% internal destruction and only 15% of viable primary production to new infection, the virus will have destructed all his alveoli around day 29, and the patient is dead. That factory closes forever. However, specific defenses against viruses tend to come up from day 7 to day 14 after the onset of infection, different in different persons, halting the exponential growth and creating degrowth for recovery. Taking day 17 as a reference emission, the airborne emissions is of the Standard Person is 74 321 virions per hour. This is the first emission flow, next spreading in the environment.

### 3.2   Mucal flows to droplets and to intestines

The 40% virions caught in mucus is mostly swallowed, in the order of 50 000 virions per hour. If passing the stomach, they can multiply in the ACE2 receptor cells abundantly present in the small intestines. SARS-2 virions are widely present in stool excrements, also after the alveolar production and measurement in nose and throat swabs has become zero, till up to 6 weeks later (He, Wang, Li, & Shi, 2020).

Mucus not swallowed is exhaled by coughing and sneezing. Vibrations and the sheer speed of air movement in coughing and sneezing, up to 14m/s (Scharfman et al., 2016) and up to 30m/s (100km/h) (Bourouiba, 2020), makes some mucus airborne, leaving by nose and mouth. Singing, speaking, and breathing do not produce virions but may also bring out virion containing droplets, from mucus or airborne from the alveoli directly. They are very small, 10 to 1µm, and evaporate to droplet nuclei at very short distance in all normal circumstances, within 40cm. The mucus exhalation will partly be in the domain of these mini droplets adding to the airborne viruses not contained in mucus, see flow *x?%* in Figure 2. These are reckoned here as part of airborne emissions as only larger droplets fall by gravity. Even if this mini-mucus-droplets amount would be a substantial part, say one third of the mucus-caught viruses coughed out, this would not increase the airborne flow substantially.

The mucus flow exhaled in bursts as larger droplets is not airborne but falls mostly within half a meter, with some parts of the bursts reaching larger distances (Bourouiba, 2020; Bourouiba, Dehandschoewercker, & Bush, 2014; Duguid, 1946; Scharfman et al., 2016). The burst is a part of the total exhalations of a person, a onetime event, or a series of such events. It is assumed here that all virions produced in one minute and as entered in not-swallowed mucus is coughed or sneezed out, in one burst per minute. The outward route by the nose does not add virions as they are produced in the alveoli. The concentration of the wetter sneeze droplets will be substantially lower, compare the figures 3 and 4 in (Scharfman et al., 2016). These bursts are quantified, see the base Table 1. On day seventeen one sneeze/cough burst contains over 800 virions.

### 3.3   Direct measurement of emissions and model outcomes: choice of Reference Person

Measuring airborne single SARS virions is a tedious affair, first done in Toronto for SARS-1 (Booth et al., 2005) and now more often with SARS-2 (Qian et al., 2020). Recently, measurements of emissions by patients have been published by an international team (Ma et al., 2020). In not severely ill patients, in quarantine and hospitalized, they found a broad range of emissions, differing for a single person and between persons, ranging from 6 000 virions per hour to 600 000 virions per hour. Taking the logarithmic mean as measurement reference, at 60 000 this level is like the Day 17 emissions of the model at 75 000 virions exhaled per hour. The 6 000 corresponds to Day 12 and 600 000 to Day 20/21. It seems that the orders of magnitude in model and measurement are not far off track, also related to the experience of the duration of pre-symptomatic and non-symptomatic spreading of viruses and the duration of the illness thereafter. These emission outcomes for Day 17 are next used in the environmental transmission model specifying infection routes and relative volumes of potential infection. A sensitivity analysis, as with a tenfold higher production and emissions, is easily performed by simple multiplication. At an emission level of Day 21, the virion factory goes in overdrive and the patient will be seriously ill, probably in hospital, with well over one square meter of his alveoli infected. The reference person for environmental analysis is the Day 17 producer of 74 321 virions per hour.

The model from which Table 1 is derived is in the Supporting Information 1 (SI1) available a www.scienceforstategies.com/SARS-Covid. The table gives an excerpt, with the Reference Person at Day 17 in red. Research by (Ivorra et al., 2020) indicates an average time lag from exposure to PCR measurable infection of around 10 days (from Figure 7). At day 10 the person would be measurable ill and at day 17 the person would approach hospitalization, see Figure 7 in (Ivorra et al., 2020).

**Table 1. Exponentially rising SARS-2 virion production in an unconstrained infected person - selected days**

| On day: | | 1 | 2 | 5 | 9 | 13 | 17 | 21 | 25 | 28 |
|---|---|---|---|---|---|---|---|---|---|---|
| Intact virions infecting alveoli, per day | | 43 | 78 | 453 | 4761 | 49975 | 524619 | 5507238 | 57812782 | 337164145 |
| Share of alveoli infected during that day | | 0.0000% | 0.0000% | 0.0001% | 0.0010% | 0.01% | 0.11% | 1.15% | 12% | 70% |
| **Cumulative share of alveoli infected** | | **0.0000%** | **0.000%** | **0.000%** | **0.002%** | **0.023%** | **0.2%** | **2.6%** | **27.1%** | **158.0%** |
| Intact virions towards airways, per day | | 245 | 441 | 2570 | 26977 | 283192 | 2972840 | 31207682 | 327605765 | 1910596823 |
| Share leaving airborne | 0.6 | | | | | | | | | |
| Share caught in mucus | 0.4 | | | | | | | | | |
| Number of airborne virions exhaled per day | | 147 | 264 | 1542 | 16186 | 169915 | 1783704 | 18724609 | 196563459 | 1146358094 |
| **Number of airborne virions exhaled per hour** | | **6** | **11** | **64** | **674** | **7080** | **74321** | **780192** | **8190144** | **47764921** |
| *According to measurements by Ma et al (2020)* | | | | | | 6000 | 60000 | 600000 | | |
| Number of virions per exhalation | | 0.00612 | 0 | 0 | 1 | 7 | 74 | 780 | 8190 | 47764.9206 |
| Number of virions caught in mucus per day | | 98 | 176 | 1028 | 10791 | 113277 | 1189136 | 12483073 | 131042306 | 764238729 |
| Share and number in mucus swallowed | 0.8 | 78 | 141 | 822 | 8633 | 90622 | 951309 | 9986458 | 104833845 | 611390983 |
| In mucus NOT swallowed | 0.2 | | 35 | 206 | 2158 | 22655 | 237827 | 2496615 | 26208461 | 152847746 |
| Virions into mucus per hour | | 4 | 7 | 43 | 450 | 4720 | 49547 | 520128 | 5460096 | 31843280 |
| **Virions produced to mucus in 1 minute** | | **0** | **0** | **1** | **7** | **79** | **826** | **8669** | **91002** | **530721** |

### 3.4   Discussion on low spreading and super spreading

The exponentially rising illness model suggests the following conclusions:

-A small exposure leads to later severity of illness, with time for virus-specific defense mechanism to develop and bend the rising curve of the person. A high exposure reduces this time.

-Well-developed generic defense mechanisms reduce the exponential rise, and weaker persons fall ill earlier and more severe.

-virions production volume is not influenced by breathing, speaking, singing, coughing, or sneezing; only the mode of emission is.

-Persons can differ very substantially in the number of virions produced and emitted.

-Most emitters are asymptomatic, and illness may occur without coughing and sneezing.

-Superspreaders may indeed emit more. However, in well documented cases the reason they infected many was primarily the highly different number of persons visiting them (Shen et al., 2004) p.258. (Shen et al., 2004) and similar (Lloyd-Smith, Schreiber, Kopp, & Getz, 2005) conclude that for many directly infectious illnesses it is environmental circumstances that may create the "20% creates 80%" rule.

## 4   METHODS 2: Exposure Models
### 4.1   Environmental routes of SARS-2 from emissions through environment to exposures

The routes to exposure cover all pathways considered in the literature. The environment characteristics have been reduced to generally occurring situations: relative humidity lower than 90% and temperature range not exceeding 30 degrees. Outside a light air of 1 meter per second, virtually no wind. Nine base routes are modeled mostly quantified.

Number 1. Is transmission in a closed space (room, ward, cabin) for a longer duration, as in visiting a bar for 2 hours or staying with an infected person for 24 hours. Number 2 covers transmission person-to-person directly, with normal exhalation at close distance in a still room. This version is generalized in 3, with transmission in a multi-person open space, with more turbulence. The bursts of sneeze and coughs are split, with number 4 covering the airborne part and number 5 the inhalation of the larger droplets. These droplets also from the basis of contaminating objects; route 6 exposure to fomites. Purely qualitative are the last three routes: the possible exposure to virions from other animals (7); from human excrements (8); and from eating food (9).

*1. Transmission in a closed space (room, ward, cabin)*

The closed space has one emitting person. Virions decay and they leave the room by ventilation. Virions which left the room don't decay there and virions decayed cannot leave anymore. For the combined effects a model has been developed as an analogon to Ohm's Law with two resistors, see SI4. The closed space is a 20m$^3$ Reference Room with a Day 17 Reference Person exhaling 74321 virions per hour (Table 1) at start and rising by 2.5% per hour (doubling time 28 hours). Normal exhalation volume and hence inhalation volume is 0.5m$^3$ per hour. The period covered is 24 hours starting with the ill person entering the virus-free room. There is no coughing and sneezing. Decay starts directly, as exponential reduction: the demise of one virion does not influence the demise of others. The median half-life of airborne SARS-2 virions is 1.1 hour with (Van Doremalen et al., 2020) Figure 2. (Schuit et al., 2020) come to 90% reduction in 286 minutes: a half-life of 1.2 hours. (Smither, Eastaugh, Findlay, & Lever, 2020) discern situations between 1- and 3-hours half-life. They all refer to mechanically aerosolized droplets under lab circumstances. (Schuit et al., 2020) measures 90% loss already at 19 minutes in winter and 8 minutes in summer, lab-simulated. (Fears et al., 2020) measured still viable SARS-2 virions in aerosols after 16 hours, which may be in line with any of the exponential reductions. A 2-hours half-life is used in modelling, with sensitivity analysis 1 and 2 hours, see table 2. Ventilation rates (VR, also: Air Change rate per Hour, ACH) vary widely. The minimum norm for Dutch schools is around VR4 (5.5liter fresh air per hour per pupil) but is not measured and enforced. In English schools, new at the time, (Coley & Beisteiner, 2002) measured VRs around 0.5. Residences and apartments are in the range of VR0.25 to VR2, with limited differences globally (Hou et al., 2019) Table 1. Norms for residential and commercial buildings are substantially higher, see Wikipedia 'Air Changes Per Hour'. Operation theaters have a minimum VR of 30. The VR0.1; VR0.5, VR2; VR5; VR10; and VR20 are used for concentration quantification. The concentration build-up in time is in Figure 3, leaving out the also very flat VR10 and VR20. Without ventilation, the half-life determines the rising equilibrium concentration. With extreme ventilation, the half-life becomes irrelevant. See Figure 3 for the concentrations resulting, and item SI2 in the Supporting Information for the basic principles, at www.scienceforstrategies.com/SARS-Covid .

**Figure 3. Virion concentration with several ventilation rates (till VR5) rising in time**

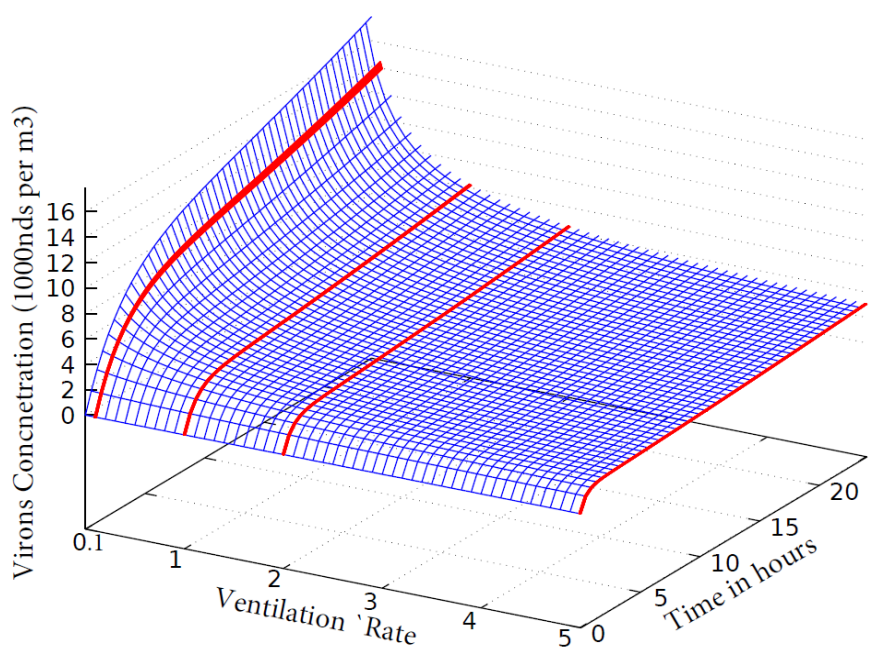

The number of virions to which persons joining in the room are exposed depends on the time of entering after the ill person and on the duration of stay. Staying with the person for 2 hours, starting at hour 0, hour 4, and at hour 12, are quantified for the exemplary VRs. Staying with an ill person in a room for 24 hours is the maximum exposure considered. See Table 2 for the outcomes (and the table covering HL1 and HL3 fully in SI2).

**Table 2 Exposure with different staying periods, Half-Lives (HL) and Ventilation Rates (VR)**

| period | HL | VR:0.1 | VR:0.5 | VR:2 | VR:5 | VR:10 | VR:20 |
|---|---|---|---|---|---|---|---|
| hrs 4 - 6 | 1 | 10,023 | 7,363 | 4,573 | 3,311 | 2,667 | 2,208 |
| hrs 0 - 2 | 2 | 5,682 | 4,801 | 3,488 | 2,696 | 2,237 | 1,887 |
| hrs 4 - 6 | 2 | 16,022 | 10,455 | 5,647 | 3,841 | 3,000 | 2,432 |
| hr5, 5 min. | 2 | 668 | 436 | 235 | 160 | 125 | 101 |
| hr5, 15 min. | 2 | 2,003 | 1,307 | 706 | 480 | 375 | 304 |
| hrs 12 - 14 | 2 | 21,653 | 13,038 | 6,890 | 4,682 | 3,657 | 2,965 |
| hrs 0 - 24 | 2 | 240,015 | 148,592 | 80,116 | 54,824 | 42,961 | 34,903 |
| hrs 4 - 6 | 3 | 19,502 | 12,072 | 6,125 | 4,058 | 3,131 | 2,518 |

The Reference Room can cover many normal situations. The number of seats in a dining room of a restaurant is typically 1 per $m^2$. With a ceiling 4m high the floor is $5m^2$, with then possibly 4 persons next to one Reference Person. With four ill persons, the fifth person would be exposed to a four times higher concentration. All are exposed equally in the well-mixed air in the room. The visitor to a bar with VR0.1 from hour 4-to-6 possibly inhales around 16 thousand virions, 20 times the infective dose of 800 virions. The start date of illness would then be around day 11 (see table 1 and SI1 in www.scienceforstrategies.com/sars-covid), with much reduced time to build up specific defenses before severe illness. In rooms with not well-mixed air the virion concentration might even be higher locally. Bad mixing occurs mostly with low VRs. An extreme case of limited mixing results from temperature inversion, quite common under bad ventilation, severest in not high rooms. The exhaled air there rises to the equilibrium mass density, possibly at standing height. That layer may have an order of magnitude higher virions concentration than well-mixed air (Zhou, Qian, Ren, Li, & Nielsen, 2017), raising the potential exposure in the low ventilation rate pub even higher.

## *2. Transmission person-to-person directly with normal exhalation*

The situation considered here is closed room with still air, hence with little or no ventilation. Normal exhalations of a healthy adult have a volume of around half a liter. Less healthy persons breath faster but with a similar volume over time. Deep inhalations and exhalations can double the tidal volume but halves the rate. Our standard person is infected but still quite healthy, on Day 17. It emits 74 virions per exhalation of the Standard Person. The literature gives a plume at exhalation of around 30 degrees, with little variation in (Gupta, Lin, & Chen, 2010) Table 2, but with more variation and directional differences shown in (Xu, Nielsen, Liu, Jensen, & Gong, 2017) and (Olmedo, Nielsen, de Adana, Grzelecki, & Jensen, 2010). The plume cone has a virtual starting point set at 5cm inside mouth or nose. The center flow may be directed downward of fully upward in lying. It tends upwards due to lower density of the warm and moist exhalation. Mixing with air reduces the speed, with air transport and turbulence taking over in spreading. The slowdown is to around 20cm/s at 40cm already, irrespective of starting speed, see Figure 6 in (Xu et al., 2017). Quantification is based on an assumed exhale sphere. At 50 cm the sphere has a volume of 13 liter, and for the hypothetical spheres at 100 and 150cm they are 93 and 300 liter. Figure 4 depicts the expansion of an exhale, moving horizontally and vertically.

The maximum exposure is in the exhalation flow directly in the face of the receiving person, who inhales in the exhalation flow fully. On average this can happen half of the time. That person inhales half of the half a liter from the 74 virions sphere of 13 liter, that is around 1.3 virions per average inhalation. At 93 and 300 liter, the hypothetical inhalation reduces to 0.20 and 0.06 virions. Person-to-person inhalations become more relevant if repeated very often, a route better approached by the closed room situation (1) or by the open space situation (3).

**Figure 4. Exhalation clouds expanding in closed spaces**

**Figure 5. Exhalation clouds expanding in open air**

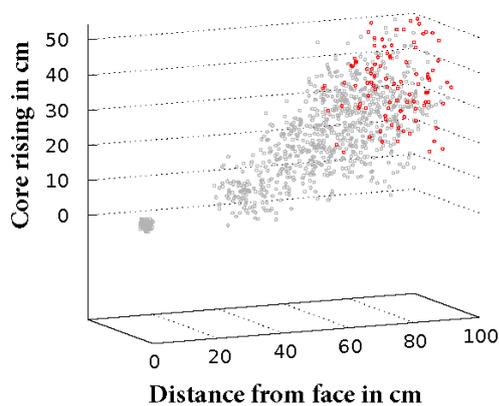
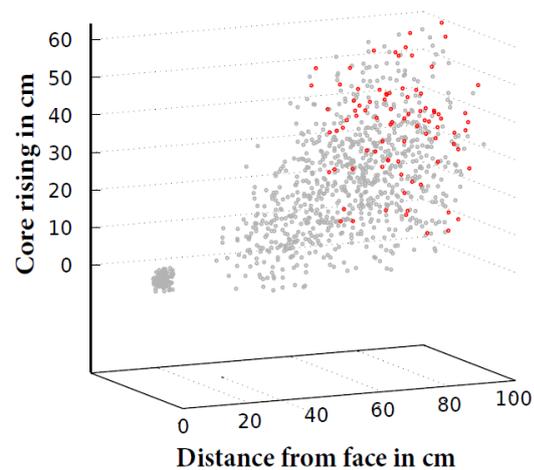

## *3. Transmission in a multi-person open space*

A standard open space is defined here as a large surface with 1 person per $m^2$ like in a well-filled terrace. With 1% ill persons there is an average distance of 10 meters between them. With 10% emitting, a high estimate at the peak of the epidemic, the average distance reduces to 3.3m. The average person will be at a safe distance, but some will stand closer and emitters may stand closer.

For inhaled concentrations, the cone of the single person-to-person still room situation, see Figure 4, is starting point, to be adapted. Open air has more turbulence, mixing faster, the cone then spreading from 30 degrees to 40 degrees, and the plume rising at 10cm/s(Olmedo Fig 3). Horizontal air movement contributes to mixing. The unusually low wind speed of 1m/s overrules the kinetics of exhalations already (wind 100cm/s versus exhale 20cm/s). Also vertical mixing, over hundreds of meters, can be substantial, especially at day time (Zhang & Rao, 1999). More local mixing is by surface heating by solar irradiation, creating convective thermals and eddies. See for this and other physical mechanisms (Vuorinen et al., 2020) p.17 and (Olmedo et al., 2010). These mechanisms together determine the speed of exhale dilution, much higher than in a room with still air. Not all open spaces are equal. A partially open football stadium, protecting viewers from rain, may function as a very large, closed room, with number of emitters, presence time, and ventilation rate to be specified.

The person-to-person inhalation situation at 50cm is uncomfortably close mostly, except for family and dear friends. Using the 40 degrees cone approach, a full inhalation at 1 meter gives a chance on inhalation of 0.08 virions, while staying in that exhalation plume for one hour, in the right direction, gives 79 virions inhaled. Staying fully in the assumed plume at 1.5m for a full hour, even more unrealistic, gives 25 virions inhaled. Reckoning with a faster rising and diluting plume in more normal outside conditions will reduce these amounts substantially, while higher winds and solar induced turbulence reduce exposure even further. The assumptions used may therefore lead to overestimation of exposures by an order of magnitude. A high estimate on potential exposure at terrace or beach is therefore set at **8 virions for 1-hour stay.**

### *4. Transmission person-to-person in burst exhalations like cough and sneeze*

Coughs and sneezes emit the virions in two ways, as a normal exhalation airborne directly, and as droplets escaping in mucus. Mini-droplets, below 5μm, evaporate near instantaneously. Larger droplets of 10μm evaporate to droplet nuclei within 0.3 second and of 100μm ( 0.1mm) within 1.3 seconds, at 50% relative humidity, see (Vuorinen et al., 2020) p.6. Droplet nuclei, below 5μm, stay airborne long-time. Coughs have far fewer droplets than the more diluted fluid sneezes, see Fig. 3 & 4 in (Scharfman et al., 2016). Sneezes also have higher speeds with a more extended vortex, highest for women having a smaller nose opening. All visualizations capture larger particles, hardly the submicron ones. The not-airborne particles are the larger droplets which fall on short distance (< 50cm) by gravity (Bourouiba et al., 2014) Fig. 5.

There is substantial discussion on 'mini-droplets'. They may form in the tubes of the alveoli and smallest bronchioli at closing after exhalation and opening at inhalation (Bake et al., 2019), directly from the alveoli, not mucus based. Or they may come indirectly from any virions caught in mucus on the epithelium in the respiratory system. The high air speed of the cough and sneeze in the higher airways may then pick them up and move them out, instead of the normal epithelial transport route (20cm per hour) to the larynx for swallowing (Sears, Yin, & Ostrowski, 2015). Measurements on numbers are lacking. In larger droplets (>5μm) 30% are infected, below that size 40% (Vuorinen et al., 2020) p.6, who also note that literature data are remarkably inconsistent. We assume here that the full 1-minute alveolar production of viable virions goes to mucus exhaled in one cough or sneeze. Part of it falls in larger droplets and the other part is transported in the vortex cloud and ultimately evaporates. We assume that 10% of mucus-based exhale falls in larger droplets and 90% leaves airborne. Other distributions are possible, without much effect on inhaled numbers, but probably with different viability.

For quantification we use the figure in the most recent publication (Bourouiba, 2020). There the burst vortex expands horizontally to 8 meter, much longer than in all previous publications, probably based on very humid and quite warm air, standing still in a lab. The quantification is based on one inhalation, full in the sneeze flow at 1-meter distance. The deep exhalation has twice the normal airborne number, that is 148 virions, while the 90% of the mucus-based airborne virions is 743, in total 893 virions. These virions are dispersed in the sneeze sphere passing at 1 meter, with a diameter of 1 meter (measured from Figure) and a volume of 3m$^3$. An extreme burst leads to fast extreme dilution. The inhalation takes half a liter from that sneeze sphere passing, that is 0.017% of 893 virions: **0.15 virions**. It seems highly improbable that a person will be exposed fully to more than one sneeze burst.

### *5. Transmission person-to-person by inhaling large droplets from exhalation bursts*

The larger droplets of cough/sneeze contain 10% of the exhaled mucus, see Route 4. By far the largest amount is at sneezing where the mucal virions are diluted by the waterier sneeze. Larger droplets have been visualized in cough and sneeze high speed pictures. Using the most detailed picture available (Bourouiba et al., 2014) Fig.5, a rough estimate of the number of droplets can be made, in the order of 50. These 50 droplets from the Standard Person then contain 74 virions, 1.5 virions per droplet.

For exposure we assume a person lying face-up on the back. As the droplet spree covers at least 1m$^2$ around the lying person and the nose openings together are not more than 5cm$^2$, the chance of being hit by the 1.5 virion droplet is 5%. Droplets falling into the mouth would be swallowed directly. The large droplet caught in the nose will be partly transported by the epithelium towards the larynx for indirect swallowing. The remainder, set at 50%, 0.75 virions, may infect the nose, where the same concentration of ACE2 receptor cells is present as in the alveoli, albeit at a very small surface. A **common cold** might develop, accompanied by anosmia in several cases (Yan, Faraji, Prajapati, Ostrander, & DeConde, 2020). There is no evidence of SARS-virion transmission from nose infection to the alveoli, which would require airborne virion production in the nose. Surface spreading is unlikely because there are hardly ACE2 receptor cells in the upper airways (Hamming et al., 2004), very different from influenza infection which spreads from the nose. When infected in then nose, the number of cells with ACE2 receptors increases fast, most probably as part of protective mechanisms against incoming viruses (Sungnak et al., 2020). As indirect evidence for an indirect pneumonia result is also lacking; the route to alveolar infection is highly improbable. The number of virions effectively reaching the alveoli is set at **zero.**

## *6. Transmission by contact with contaminated objects: fomites*

Droplets, especially larger ones, can fall on persons and objects, where they can remain intact for longer periods, especially on very smooth surfaces such as polished stainless steel, see (Van Doremalen et al., 2020) Table 1 and see the later survey by (Aboubakr, Sharafeldin, & Goyal, 2020). Viable SARS-2 virions remained detectable for up to several days, after exposure with larger amounts of lab produced viruses. Real life measurements in hospitals with SARS-1 patients showed extensive spread of virions, measured by PCR test, see (Dowell et al., 2004). However, no viable viruses could be gathered. First order modelling gives a similar answer. Assume that two larger cough or sneeze droplets of the total of 50 droplets have fallen at a certain smooth surface and stayed there, in total 3 virions (2/50*74). Before their demise, one finger wipes them off, a main part of them, 2 virions. With a less smooth surface that is not possible. That finger is brought to the mouth or nose. Mouth infections have not been reported. Destruction and swallowing with saliva may be expected. With a finger containing 2 virions, half of them might be smeared into the nose opening, **1 virion**, a high estimate. Nose-epithelium will transport most virions to the larynx for swallowing, but not this one, then remaining available for local infection. A nose infection is possible if that one virion overcomes the normal defenses against infection, which is extremely improbable. The net direct intake to the alveoli for pneumonia is set at **zero.**

## *7. Transmission by human excrements*

Infected excrements have widely been documented, see (Wang et al., 2020) and (Heneghan, Spencer, Brassey, & Jefferson, 2020). Stool can be infected from production of virions in the small intestines, the second largest potential production site after the alveoli. The intestines can produce virions after the last positive test of respiratory samples, up to well over a month (Wu et al., 2020), and have shown to be viable for infection in SARS1 and MERS. Stool PCR tests are equally usable for checking on a person's infection as nasal swabs (JingCheng Zhang, Wang, & Xue, 2020). This orofecal route has not been substantiated for SARS-2 but remains a serious possibility, so they state. The intestine infection can be by the oral route or by transport from the alveoli in the blood stream, or the other way around.

The amount and period of infection at a local population level can be indicated by the number of SARS-2 virion parts in the sewer system measured by PCR, see (Medema, Heijnen, Elsinga, Italiaander, & Brouwer, 2020). These are assumed not to be viable for infection anymore. Stool can spread to mouth and nose of other persons, especially in care situations. Stool might also be aerosolized in flushing toilets (Johnson, Lynch, Marshall, Mead, & Hirst, 2013), but has not been documented in relation to SARS-2. However, in several instances in hospitals higher concentrations of airborne SARS-2 has been measured in toilet rooms.

Cases of intestine infection leading to a new alveoli infection, with pneumonia, have not been documented or indicated. Direct intestine infection has been documented in MERS, see (Shi, Sun, & Hu, 2020) p.942. The primary intestine infection might be a serious issue but as yet not substantiated route, signaled as (**!!!!**) in Table 3.

### *8. Transmission from other animal species to persons*

Zoonotic transmission forms the basis for the new SARS-2 infection in humans, with the ultimate reservoir in bats. The spread is however mainly human to human, as in SARS-1. In MERS the spread was also between camels and humans. Animals with ACE2 receptors and mutual infections are several, see the survey in (Opriessnig & Huang, 2020). Living in close vicinity with humans are cats and ferrets, as quite popular pets and as farmed for their fur. They have been infected by humans and have symptoms like humans. There are no known cases of cat to human infection yet, but close contact for a longer duration is not uncommon. There is some slight evidence that cat fleas can carry the virus, and visit other animals (Villar et al., 2020). Farmed ferrets have been infected by their caretakers with human SARS-2 and large-scale mutual infection has been established in ferret farms in the Netherlands (production 5 million per year). There is no research yet on inhalation risks of keeping them as pets, which is widespread in the US, third place after dogs and cats. Dogs are less prone to infection, while pigs, chickens and ducks are not prone to infection (Jianzhong Shi et al., 2020). There are many non-household animals prone to infection, with armadillos, Syrian hamsters and rhesus macaques used in experiments as examples (Shan et al., 2020).

Though of some concern, the risk of exposure from these sources seems very low, with a possible exception for ferrets kept as pets; warning (**!!**).

### *9. Transmission by food*

Many wild animals are eaten, including those that can carry SARS-2, with armadillos as an example (Shan et al., 2020), and some are raised for eating, as dogs in many countries. Also ferrets after having been cleaned of their fur are animal fodder, possibly ending up higher in the food chain. Another route is by contamination of food, by human droplets and stool and stool from any SARS-2 infected animal, such as cats, ferrets, and several other rodents. SARS-2 virions have been found on frozen food, where they might remain long time viable, till thawing. There then is an exposure route to intestine infection by eating raw food, and a route to nose infection by handlers of the food as a fomite. The number of virions involved might be much larger than with objects contaminated by cough and sneeze droplets. There are no documented cases yet where the SARS virus has been transmitted by food. The risk of SARS-2 pneumonia by consuming food containing virions seems therefore low, and then might lead to a primary intestinal infection. The risks will be highest with persons handling the contaminated food very regularly or eating just unfrozen food raw. Developing SARS-2 pneumonia from food seems extremely improbable, flagged but set at **Zero**.

## 5   RESULTS 2: Potential exposures
### 5.1   Routes to exposure compared

Table 3 integrates exposures results (see the excel files in SI3 at www.scienceforstrategies.com/SARS-Covid ). The relative risks of the routes to exposure differ by six orders of magnitude. Staying in the standard room with an ill person for two hours or longer always poses a serious risk, even with good ventilation (VR5), which is not reached in many current situations, not even in many schools, see (Coley & Beisteiner, 2002). Orders of magnitude lower are risks for shorter stays and close contacts with ill persons inhaling their exhales for a serious duration, specified in controlled lab circumstances.

Single events, like short-distance face-to-face breathing and sneezing, and fomites infection link to an exposure risk several orders of magnitude lower again.

**Table 3. Exposure routes ordered as to potential virions intake**

|  | Virions: | Normalized to Room exposure hr 4-6, VR5 |  |
|---|---|---|---|
| Room exposure* hr 0 - 24 VR0.1 | 240015 | 62 | Staying in hospital ward very bad ventilation |
| Room exposure* hr 0 - 24 VR0.5 | 148592 | 39 | Staying in hospital ward bad ventilation |
| Room exposure* hr 0 - 24 VR2 | 80116 | 21 | Staying in hospital ward reasonable ventilation |
| Room exposure* hr 0 - 24 VR5 (HL3) | 57870 | 15 | HL3, Staying in hospital ward normal ventilation |
| Room exposure* hr 0 - 24 VR5 | 54824 | 14 | Staying in hospital ward normal ventilation |
| Room exposure* hr 0 - 24 VR5 (HL1) | 47349 | 12 | HL1, Staying in hospital ward normal ventilation |
| Room exposure* hr 0 - 24 VR10 | 42961 | 11 | Staying in hospital ward good ventilation |
| Room exposure* hr 0 - 24 VR20 | 34903 | 9 | Staying in hospital ward very good ventilation |
| Room exposure* hr 12- 14 VR0.1 | 21653 | 6 | Visiting ill family member very bad ventilation |
| Room exposure* hr 4 - 6 VR0.1 | 16022 | 4 | Visiting ill family member very bad ventilation |
| Room exposure* hr 12- 14 VR0.5 | 13038 | 3 | Visiting ill family member bad ventilation |
| Room exposure* hr 4 - 6 VR0.5 | 10455 | 3 | Visiting ill family member bad ventilation |
| Room exposure* hr 12- 14 VR2 | 6890 | 2 | Visiting ill family member reasonable ventilation |
| Room exposure* hr 0 - 2 VR0.1 | 5682 | 1 | Visiting bar very bad ventilation |
| Room exposure* hr 4 - 6 VR2 | 5647 | 1 | Visiting ill family member reasonable ventilation |
| Room exposure* hr 0 - 2 VR0.5 | 4801 | 1 | Visiting bar bad ventilation |
| Room exposure* hr 12- 14 VR5 | 4682 | 1 | Visiting ill family member normal ventilation |
| **Room exposure* hr 4 - 6 VR5** | **3841** | **1.000** | Visiting ill family member normal ventilation |
| Room exposure* hr 12- 14 VR10 | 3657 | 0.952 | Visiting ill family member good ventilation |
| Room exposure* hr 0 - 2 VR2 | 3488 | 0.908 | Visiting bar reasonable ventilation |
| Room exposure* hr 4 - 6 VR10 | 3000 | 0.781 | Visiting ill family member good ventilation |
| Room exposure* hr 12- 14 VR20 | 2965 | 0.772 | Visiting ill family member very good ventilation |
| Room exposure* hr 0 - 2 VR5 | 2696 | 0.702 | Visiting bar normal ventilation |
| Room exposure* hr 4 - 6 VR20 | 2432 | 0.633 | Visiting ill family member very good ventilation |
| Room exposure* hr 0 - 2 VR10 | 2237 | 0.582 | Visiting bar good ventilation |
| Room exposure* hr 0 - 2 VR20 | 1887 | 0.491 | Visiting bar very good ventilation |
| Still room face-to-face 1h3 50cm | 1309 | 0.341 | 1000 inhalations at 50cm |
| Room exposure* hr 5, 5 min. VR0.1 | 668 | 0.174 | Visiting ill family member very bad ventilation |
| Room exposure* hr hr 5, 15 min. VR5 | 337 | 0.125 | Visiting ill family member normal ventilation |
| Still room face-to-face 1hr 100cm | 201 | 0.052 | 1000 inhalations at 100cm |
| Room exposure* at hr 5, 5 min.VR5 | 160 | 0.042 | Visiting ill family member normal ventilation |
| Open air face-to-face1hr 100cm | 79 | 0.021 | 1000 inhalations at 100cm |
| Still room face-to-face 1hr 150cm | 60 | 0.016 | 1000 inhalations at 150cm |
| Open air face-to-face1hr 150cm | 25 | 0.006 | 1000 inhalations at 150cm |
| Still room face-to-face 1inh 50cm | 1.31 | 0.00034 | 1 inhale at 50cm |
| Animals to humans | 1 | 0.00026 | (!!) Serious option, not yet measured (ferrets, cats) |
| Still room face-to-face 1inh 100cm | 0.20 | 0.00005 | 1 inhale at 100cm |
| Burst inhalation at 100cm | 0.150 | 0.00004 | Extreme sneeze burst (Bourouiba, 2020) |
| Open air face-to-face 1inh. | 0.079 | 0.00002 | 1 inhale at 100cm |
| Still room face-to-face 1inh. 150cm | 0.060 | 0.00002 | 1 inhale at 150cm |
| Open air face-to-face 1inh. 150cm | 0.025 | 0.00001 | 1 inhale at 150cm |
| Burst droplet exposure | 0 | 0.00000 | Extreme burst (Bourouiba, 2020) |
| Fomite into nose | 0 | 0.00000 | Possible nose infection by hand to nose |
| Human excrements | 0 | 0.00000 | (!!!!) Substantial intake to nose and intestines (carers) |
| Food | 0 | 0.00000 | Nose infection (workers) very low chance intestines |

*All room situations are with Half-Life 2 hours, except where indicated differently.

# 6    DISCUSSION: Models and Exposures

## 6.1    Restrictions in models and data.

Several shortcomings are present now, and only partly repairable.

The exponential infection model in Methods 1 does not reckon with persons with lower or higher non-specific defense mechanisms as related to higher or lower speed to illness. Sensitivity analysis would be useful as data are lacking. Indirect information may come up in the coming months and years.

The closed space model has been developed after Ohm's Law. Better models might be developed or adapted from other domains.

The exhalation–inhalation models follow the cones approach from the literature for quantification, with spheres growing when expanding through the cone, and the cone opening increasing from 30 degrees to 40 degrees with turbulence. These models seem overly simplistic and empirically linked to lab situations only.

The most recent empirical sneeze model (Bourouiba, 2020) forms the basis for sneeze quantifications, at 1-meter distance and for droplets, but might be related to special circumstances like extreme still-standing warm and humid air.

Real-life quantified measurements of viable virions will remain extremely burdensome and costly.

Therefore, quantified measurement of actual virions in flows beyond hospitals and labs is lacking, and key variables like virions half-life in realistic airborne and fomite situations are lacking as well.

Routes from fomites to actual intake to mouth and nose have never been measured, making claims on this route yet ill-supported.

## 6.2    From standard persons and standard situations to more realistic versions

Resolving these restrictions for more adequate and realistic modelling will take time. For now, better sensitivity analysis on key variables is a main option.

Sensitivity analysis for choice of Standard Person Day 17 will increase all exposures proportionally but with all relative scores the same. Wide variations would bring the dynamic model in a highly improbable domain, reaching death much sooner or hardly at all.

Changing the distribution of primary virion production over the airborne and mucal outflows will also leave the relative scores mainly intact.

The closed room routes did not cover ventilation air moved to other rooms as a source of exposure, widely discussed and often inferred from infections (see SI6). A most serious case would be a fresh air ventilation rate of only 50%, quite common, with the other half coming from the closed room cases. Then concentrations in the next room would be half of the closed room case, in many cases still serious.

The fomite infection route discussed indicates that few virions can effectively be brought into the nose at most. Real-life experiments could show some evidence, as is present for the very different influenza virus. Could such routes be established for SARS-2?

## 6.3    From exposure to infection

Exposure to virions has been set equal for all forms. In terms of infectiveness there are very large differences, however. Medicine research (Brown et al., 2002) has shown that finest particles reach deeper into the lungs than fine particles, UFPs against PM2.5 in air pollution language. Single SARS virions, 120 nanometer diameter, are UFPs. Evaporated droplets, droplet nuclei, are up to 5μm, larger even than PM2.5, while the maximum particle size reaching alveoli is 3μm, see (Talaat & Xi, 2017), especially p.20 and Figure 12. Inhaled medicines come deeper the smaller their size is, with the share of UFPs highest, again the smallest coming deepest. So, it is not just the number of virions exposed to but also their size in determining the risk on SARS-2 pneumonia. Larger droplet nuclei will play a smaller role per virion than single virions as they are caught more in the upper airways, see also (Mittal et al., 2020) p.894. The number of virions per mini-droplet and droplet nucleus is not known in relation to their size, nor is their viability. Here they have been counted equal. Transport models developed for medicine application could improve infection modelling.

Infection of alveoli requires virions to pass the tracks, never established for SARS-2, as by contrast for influenza. The flu spreads in these tracks and destructs the protective layers there, including by macrophages (Ghoneim, Thomas, & McCullers, 2013). The pneumonia ultimately resulting is not viral as with COVID-19, but bacterial, mostly Streptococcus Pneumoniae. Could the viral infection routes be established for the different forms of SARS-2?

### 6.4    From analysis to technical and policy measures

Exposure prevention measures have not been investigated here, nor have medicines. Ill ventilated spaces are of prime concern, especially where several ill and highly emitting persons could be present. Options differ deeply between personal short-term measures like good masks in badly ventilated spaces, to long-term collective measures such as effective standards for virus-free ventilation in all public buildings and transport spaces, and also in private buildings.

## 7    Conclusions: SARS-2 Risks

1. The outcomes indicate clearly where measures may be more useful, much less so, or not at all.
2. All long stay situations in closed rooms, also with reasonable to good ventilation VR2 to VR5, pose a risk of exposure, by up to well over 10,000 virions. This seems enough for severe infection.
3. Incidental and short period exposures are well below the 1000 virions boundary, referring to extreme maximum intakes only: in still, warm and humid air inhaling the exhale of ill persons face-to-face many times.
4. Keeping 150cm distance in a well-mixed room does not protect against risky exposure, except with high ventilation rates, where distance does not matter much.
5. All outside exposures and all contact exposures, as with fomites and large droplets, lead to negligible potential exposure, including extreme sneeze bursts fully inhaled at 1-meter distance.
6. Norms on ventilation rates, often too low, differ an order of magnitude from actual ventilation rates, which can result in extreme concentration, exposure, and risk differences, especially in the range of lower VRs.
7. The SARS-2 models developed here show the strength of Industrial Ecology to engage in mass balancing beyond industrial and natural systems. Detailed medical and epidemiological knowledge and models may be linked to them by others for a more integrated view.